\documentclass[12pt]{article}
\usepackage{hpatex,epsf}

\def\gtrsim{{\scriptscriptstyle ^{\displaystyle >}_{\displaystyle \sim}}}
\begin{document}
\hfill ETH-TH/96-50\\
\vspace{-1cm}
\title{Deviations from
the Harrison-Zel'dovich spectrum due to the Quark-Gluon to
Hadron Transition}

\authors{Christoph Schmid\footnote{Seminar given at Journ\'ees Relativistes '96,
Ascona (Switzerland).}, Dominik J. Schwarz, and Peter Widerin}

\address{Institut f\"ur 
Theoretische Physik, ETH-H\"onggerberg, CH-8093 Z\"urich}

\abstract{
We investigate the effect of the quark-gluon to hadron transition 
on the evolution of cosmological perturbations.
If the phase transition is first order, the sound speed vanishes
during the transition, and density perturbations fall freely.
The primordial Harrison-Zel'dovich 
density fluctuations for scales below the Hubble radius at the transition
develop peaks, which grow linearly with the wavenumber,
both for the hadron-photon-lepton fluid and for cold dark matter.
The large peaks in the spectrum produce cold dark matter clumps
of $10^{-8}$ to $10^{-11} M_\odot$.
}

QCD makes a transition from a quark-gluon plasma
at high temperatures to a hadron gas at low temperatures. 
Lattice QCD simulations give a transition temperature 
$T_\star \sim 150$ MeV and indicate a 
first-order phase transition for the physical values of the u,d,s-quark masses
\cite{lattice}. The relevance of the QCD transition for cosmology,
especially for big-bang nucleosynthesis \cite{BBN}, has been discussed
before, but the focus was on effects of bubble formation. 
In this paper and in \cite{SSW} 
we look at matter averaged over scales 
much larger than the bubble separation. We
show that for a first order phase transition the sound speed $c_s
= (\partial p/\partial \rho)_s^{1/2}$ drops to zero suddenly at the moment
the transition temperature $T_\star$ is reached, stays zero for the entire 
time until the phase transition is completed, and
afterwards suddenly rises back to $c_s \approx
c/\sqrt{3}$. In contrast the pressure
stays positive and varies continuously, although it goes below the radiation
fluid value $p/\rho = 1/3$. Since $c_s$ is zero
during the transition, there are no pressure perturbations, no pressure 
gradients, no restoring forces.
Pre-existing cosmological perturbations, generated by
inflation with a Harrison-Zel'dovich spectrum,
go into free fall for about a Hubble time. The superhorizon modes 
(at the time of the transition) remain unaffected, the subhorizon modes 
develop peaks in $\delta\!\rho/\rho$ which grow linearly with wavenumber,
$\sim k/k_\star$, where $k_\star^{\rm phys} \sim$ Hubble rate $H$ at the end
of the QCD transition.

The sound speed, $c_s = (\partial p/\partial \rho)_s^{1/2}$, must be zero 
during a first-order phase transition of a fluid with negligible chemical 
potential, since the fluid must obey
\begin{equation}
\label{2nd}
\rho + p = T {dp\over dT} \ ,
\end{equation}
according to the second law of thermodynamics. Because the energy
density $\rho$ is discontinuous in temperature at $T_\star$ for a first-order
phase transition, the pressure $p$ must be continuous with
a discontinuous slope. As the universe expands at fixed temperature $T_\star$
during the phase transition, $\rho$ as a function of time slowly decreases
from $\rho_+(T_\star)$ to $\rho_-(T_\star)$, $p$ stays constant at 
$p(T_\star)$, and therefore $c_s$ is zero during the whole time of the
phase transition.

The interaction rates in the QCD-photon-lepton fluid
are much larger than the Hubble rate, $\Gamma/H\gg 1$,
therefore we are very close to thermal and chemical equilibrium, 
the QCD transition
is very close to a reversible thermodynamic transformation.
Estimates show that supercooling, hence entropy production,
is negligible, $(T_\star - T_{\rm supercooling})/T_\star
\sim 10^{-3}$ \cite{supercool}.
Bubble formation is unimportant for our
analysis, estimates give a bubble separation $\ell_b \sim 1$ cm
\cite{Christiansen}, while the Hubble radius at the QCD transition
is $R_H \sim 10$ km. We shall analyze perturbations with $\lambda \gg \ell_b$.

The bag model gives a parameterization
and a reasonable fit to the lattice QCD data \cite{DeGrand}. 
In the bag model it is assumed that for $T>T_\star$ the quark-gluon plasma
(QGP) obeys
\begin{equation}
\label{p}
p_{\rm QGP}(T) = p_{\rm QGP}^{\rm ideal}(T) - B \ ,
\end{equation}
where $p_{\rm QGP}^{\rm ideal}(T) = (\pi^2/90) g_{\rm QGP}^* T^4$, $g^*$ is the
effective number of relativistic helicity states, and $B$
is the bag constant. We include u,d-quarks and gluons in the 
quark-gluon plasma, 
and for $T<T_\star$ we have a hadron gas (HG) of pions, which 
we treat as massless and ideal.
$\rho$ follows from 
Eq.~(\ref{p}) via the second law, Eq.~(\ref{2nd}), and 
$s$ from $s = dp/dT$. 
The bag constant is determined by the critical temperature $T_\star$ via
$p_{\rm QGP}(T_\star) = p_{\rm HG}(T_\star)$.
The photon-lepton fluid contains $\gamma,e,\mu$, and 3 neutrinos. 
The growth of the scale factor during the QCD 
transition, $a_+/a_- \approx 1.4$, 
follows from the conservation of entropy in a comoving 
volume. 

The evolution of linear cosmological perturbations through 
the QCD transition is
analyzed in the longitudinal sector
(density perturbations) for perfect fluids. We choose a slicing 
$\Sigma$ of space-time with unperturbed mean extrinsic curvature,
$\delta[\mbox{tr} K_{ij}(\Sigma)] = 0$. 
The adapted gauge is the 
uniform expansion (Hubble) gauge \cite{Bardeen}.
As fundamental evolution equations for each fluid we have $\nabla_{\mu}
T^{\mu\nu} = 0$, i.e. 
the continuity equation and (in the longitudinal sector)
the 3-divergence of the Euler equation
of general relativity,
\begin{eqnarray}
\label{C}
\partial_t \epsilon &=& - 3 H(\epsilon + \pi) - \bigtriangleup \psi -
3 H (\rho + p)\alpha \\
\label{E}
\partial_t \psi &=& - 3 H \psi - \pi - (\rho + p)\alpha \ ,
\end{eqnarray}
where $\epsilon \equiv \delta\!\rho$, $\pi \equiv \delta\! p$,
$\rho \equiv \rho_0$, $p \equiv p_0$, 
$\vec{\nabla} \psi =$ momentum density,
$\alpha =$ lapse function. The system of dynamical equations is
closed by Einstein's $R_{\hat{0}\hat{0}}$-equation, the 
general relativistic Poisson equation,
\begin{equation}
\label{R}
(\bigtriangleup + 3 \dot{H})\alpha = 4\pi G (\epsilon + 3 \pi) \ ,
\end{equation}
together with the equation of state.
These three equations are
the Jeans equations extended to general relativity in the 
longitudinal sector. 

%%%%%%%%%%%%%%%%%%%%%%%%%%%%%%%%%%%%%%%%%%%%%%%%%%%%%%%%%%%%%%%
\begin{figure}
\begin{center}
\epsfig{figure=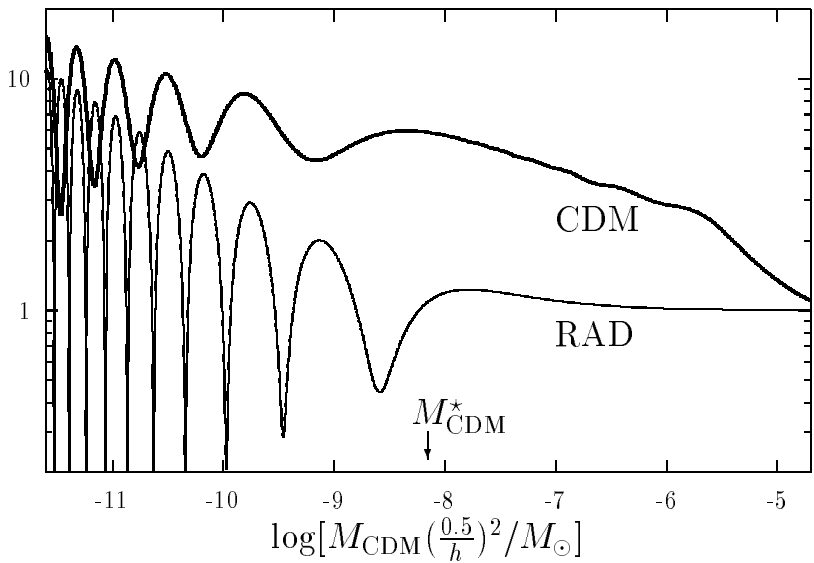,width=0.65\linewidth}
\end{center}
\vspace{-18pt}
\caption{
The modifications of the CDM density contrast $A_{\rm CDM} \equiv |
\delta_{\rm CDM}|$ at $T_\star /10$ and
of the radiation fluid amplitude $A_{\rm RAD} \equiv (\delta_{\rm RAD}^2 + 3
\hat{\psi}_{\rm RAD}^2)^{1/2}$  due to the QCD transition.
Both quantities are 
normalized to the pure Harrison-Zel'dovich radiation amplitude. On the 
horizontal axis the wavenumber $k$ is represented by the CDM mass contained
in a sphere of radius $\pi/k$.} 
\label{fig1}
\end{figure}
%%%%%%%%%%%%%%%%%%%%%%%%%%%%%%%%%%%%%%%%%%%%%%%%%%%%%%%%%%%%%%%
Our numerical results of evolving a mode $k$ of a 
cosmological perturbation through the 
QCD transition are given in Fig.~\ref{fig1}. 
We work with the dimensionless variables $\delta \equiv
\epsilon/\rho$ (density contrast) and 
$\hat{\psi} \equiv k^{\rm phys} \psi/\rho$
($\sim$ peculiar velocity).
We show the 
enhancement of the amplitude $A_{\rm RAD} \equiv (\delta_{\rm RAD}^2 + 3
\hat{\psi}_{\rm RAD}^2)^{1/2}$ of the acoustic oscillations of the
radiation fluid  
(QCD, photons, leptons) after the transition compared 
to the amplitude without transition. For cold dark matter (CDM) 
we show the amplitude $A_{\rm CDM} \equiv |\delta_{\rm CDM}|$ 
at $T_\star/10$ compared to $A_{\rm RAD}$ without transition. 
In both cases we obtain
peaks over the Harrison-Zel'dovich spectrum of adiabatic
density fluctuations.
Only those modes are affected which are subhorizon at the end of 
the transition,
$k \gtrsim k_\star$, where $k_\star^{\rm phys}(t_+) \sim H(t_+)$.
Our peaks grow linearly in $k$ for $k \gg k_\star$.
The modes $k$ are labeled
by the CDM mass contained in a sphere of radius $\lambda/2 = \pi/k$.
The value $k_\star$ corresponds to $M^\star_{\rm CDM} \sim 10^{-8} M_\odot$.
The radiation energy inside $\lambda_\star/2$ is $\sim 1 M_\odot$,
but it gets redshifted as 
$M_{\rm RAD}(a) \sim (a_{\rm equality}/a) M_{\rm CDM}$.
The perturbations in the radiation fluid will be wiped out 
by collisional damping from neutrinos on scales much smaller than the horizon 
scale at neutrino decoupling ($1$ MeV).

For cold dark matter we consider the lightest supersymmetric particle (LSP), 
the neutralino $\widetilde{\chi}_1^0$ in the minimal supersymmetric 
standard model with universal supersymmetry breaking at the grand unified 
scale. LEP 1.5 combined with the gluino mass limit from Fermilab gives
a minimal mass $M^{\rm min}_{\rm LSP} = 27$ GeV \cite{LEP}. 
With a freeze-out temperature of $T_f \sim
M_{\rm LSP}/20$ free-streaming wipes out CDM structure for $k/k_\star
> 10$.

The origin and magnitude of these impressive peaks for $k \gg k_\star$
is easily understood.
The
radiation fluid in each subhorizon 
mode makes standing acoustic oscillations 
before and after
the QCD transition with gravity negligible and with equal
amplitudes of $\delta$ and $\sqrt{3}\hat{\psi}$.
During the transition the sound speed is zero, there are
no restoring forces from pressure gradients, the radiation fluid  
goes into free fall. 
If the transition time is short, $\Delta t < H^{-1}$, 
gravity is again negligible for the radiation fluid 
during this free fall. 
This is inertial motion in the sense of Newton. 
The peculiar velocity is constant, and the density
contrast grows linearly in time with a slope $k$.
Therefore the height of the peaks is 
$(A_+/A_-)_{\rm peaks} \approx k/k_\star$.
Hence modes with $k^{\rm phys}/H \gtrsim 10^4$ go nonlinear by the end of 
the QCD transition.
The explicit solution along these lines is presented in \cite{SSW}.
CDM falls into the gravity wells generated by the radiation fluid. 

The peaks above the Harrison-Zel'dovich spectrum lead to
CDM clumps with $10^{-8} > M_{\rm CDM}/M_\odot
> 10^{-11}$, which go nonlinear sometime after equality and 
virialize by violent gravitational
relaxation. The mass range of these CDM clumps lies just beyond the
smallest mass ($5 \times 10^{-8} M_\odot$) 
accessible to present microlensing observations \cite{EROS}, thus future
lensing observations might discover such clumps.

We like to thank J. A. Bardeen, P. Jetzer, 
F. Karsch, V. Mukhanov, J. Silk, and
N. Straumann for helpful discussions.
D.S. and P. W. thank the Swiss National Science Foundation for financial
support.


\begin{thebibliography}{99}
\bibitem{lattice}C. DeTar, Nucl. Phys. B (Proc. Suppl.) {\bf 42},
         73 (1995); T. Blum et al., Phys. Rev. D {\bf 51}, 5153 (1995);
         Y. Iwasaki et al., hep-lat/9505017 (1995).
\bibitem{BBN}R. A. Malaney and G. J. Mathews, Phys. Rep. {\bf 229}, 145 (1993).
\bibitem{SSW}C. Schmid, D. J. Schwarz, and P. Widerin, astro-ph/9606125 (1996).
\bibitem{supercool}Based on G. M. Fuller, G. J. Mathews, and C. R.
         Alcock, Phys. Rev. D {\bf 37}, 1380 (1988).
\bibitem{Christiansen}M. B. Christiansen and J. Madsen, Phys. Rev. D 
         {\bf 53}, 5446 (1996).
\bibitem{DeGrand}T. DeGrand and K. Kajantie, Phys. Lett. {\bf 147B},
         273 (1984).
\bibitem{Bardeen}J. A. Bardeen, Phys. Rev. D {\bf 22}, 1882 (1980);
         and in {\em Particle Physics and Cosmology}, ed.
         A. Zee (Gordon and Breach, New York, 1989).
\bibitem{LEP}ALEPH Collaboration, D. Buskulic et al.,
         hep-ex/9607009 (1996); J. Ellis et al., hep-ph/9607292 (1996).
\bibitem{EROS}EROS Collaboration. E. Aubourg et al., astro-ph/9503021 (1995).
\end{thebibliography}
\end{document}